\documentclass[11pt]{article}

\usepackage{epsfig}
\newcommand{\be}{\begin{equation}}
\newcommand{\ee}{\end{equation}}
\newcommand{\bea}{\begin{eqnarray}}
\newcommand{\eea}{\end{eqnarray}}
\newcommand{\nn}{\nonumber \\}

\newcommand{\ba}{\begin{array}}
\newcommand{\ea}{\end{array}}
\newcommand{\beas}{\begin{eqnarray*}}
\newcommand{\eeas}{\end{eqnarray*}}

\advance \topmargin by -\headheight
\advance \topmargin by -\headsep
\evensidemargin \oddsidemargin
\advance\hoffset by -3mm  
\advance\voffset by  8mm  

\font\mybb=msbm10 at 10pt
\def\Hf {\hat{f}}
\def\Hg {\hat{g}}
\def\Hl {\hat{l}}
\def\HA {\hat{A}}
\def\bb#1{\hbox{\mybb#1}}

\def\bR {\bb{R}}

\def\bC {\bb{C}}

\def\pa{\partial}

\begin{document}

\begin{titlepage}
  \pagestyle{plain}
  \begin{flushright}
  \footnotesize
  DAMTP-2000-56\\
  {\bf hep-th/0006090}\\
  \normalsize
  \end{flushright}

  \vskip 30pt

  \begin{center}
  {\Large Geometry of the Nonabelian DBI Dyonic Instanton }\\
    \vskip 25pt

  {\large {M. Zamaklar \\
CMS, University of Cambridge, Wilberforce Road,\\
Cambridge CB3 0AW,  UK
\footnote{email: mz212@damtp.cam.ac.uk} }}
\end{center}
  \vskip 28pt

\begin{abstract}
\baselineskip=15pt 
We introduce a calculus of the Lie algebra valued functions present 
in Tseytlin's proposal for the nonabelian DBI action, and apply it to
show that the recently found dyonic instanton is a solution of the full
nonabelian DBI action. The geometry of this solution exhibits a new effect of ``blowing-up'' of the brane, which was not 
present in the case of the brane realisation of monopoles. We
interpret this solution as the superposition of the  $D0$ brane and fundamental string $F1$ which connects two separated $D4$ branes.  Both $F1$ and $D0$ are delocalised  in the ``blown-up'' region
between two separated $D4$ branes.

\end{abstract}

 \end{titlepage}

\newpage

\baselineskip=15pt

\section{Introduction}
\vskip10pt

It is now well understood that the low energy dynamics of a single
$Dp$-brane is described by the Dirac-Born-Infeld action, combined with
a Wess-Zumino term which describes the coupling of the brane to the
background RR fields and gravity curvature. This action is consistent with  T-duality, in the
sense that it can be obtained from the ten dimensional BI action by
dimensional reduction. It admits an extension to
the $\kappa$--invariant action for both flat \cite{agan} and general
\cite{ceder,ceder1,paul} type II supergravity backgrounds. This
action is effective since it describes only the dynamics of the
massless excitations of the brane. It is low energy in the sense that it
is valid only for slowly varying gauge fields, i.e. there are 
corrections to this action of order ${\rm O} ( l_{s}^{3} \partial
F)$. However, the abelian DBI action incorporates corrections to all orders in the field strength. 
The physical degrees of freedom of the DBI action are manifest in the static gauge and
they correspond to the degrees of freedom of a  $p+1$-dimensional $U(1)$ gauge theory, $9-p$ transverse scalar
fields and their fermionic partners.

On the other hand, the low energy dynamics of a  system of $Dp$ branes
is still not well understood. The lowest energy description for a
system of $Dp$-branes is given by the maximally supersymmetric $U(N)$
Yang-Mills theory in $p+1$ dimensions. But the full effective action is
unknown. One of the main problems is the following. In the
abelian case, one can uniquely split the abelian action into those parts
which do and those parts which do not depend on
the derivative of $F$. For the system of branes  that is
not the case  since $[D_{a},D_{b}] F_{cd} = [F_{ab}, F_{cd}]$. Hence, one is allowed to ``trade'' $DF$ terms for commutators of $F$ terms. To overcome this ambiguity,  when talking about the  nonabelian Born-Infeld  action (hereafter referred to as NBI action) one means only the part of the effective action which does not contain any
commutators of $F$. Tseytlin has  {\it defined} the NBI action as
that part of the full open string effective action which is completely symmetric in
all factors of $F$ in each monomial  $\rm{tr}(F...F)$. The {\it full} open
string effective action of the $Dp$-brane is  then given by the sum of the following terms: the  NDBI
action, the $F^{n}$ terms containing factors of commutators of $F$, and
the terms with symmetrised covariant derivatives. 
Hence a discrepancy between the predictions of the full string theory and
the predictions of the nonabelian DBI action (such as found in \cite{wash,bain}) is not
unexpected.\footnote{Supersymmetrisation of this action is also a nontrivial
problem \cite{tsey}.}

We will not consider here questions about the validity of
the NDBI action and proposals for possible correction terms which should be
added to this action. Instead, we will focus on rules for efficient calculation
with the existing NDBI action.
As given, Tseytlin's proposal for the nonabelian action allows one to 
work only in a power expansion of the field strength $F$. 
However, one would wish to introduce a closed-form notation,
which would allow one to do all calculations without writing explicit power
expansions. Of course, the new objects that we define are such that at each
stage of the calculation one can rewrite all results in a power expansion. 
In the literature many calculations have been done  in closed form,
implicitly introducing Lie algebra valued functions. 
However, strict definitions as well as precise rules for how one should do
calculations with these functions have not been given. The first part of
this paper aims to fill this gap.
We also point
out the limitations of this approach. Namely,
after certain operations one loses the ability  to further express things in a closed form. 

In the second part of the paper we apply these rules to find a
particular solution of the NDBI action.
It was recently found \cite{LT} that five dimensional $N=2$
\footnote{We use language of the five dimensional spinors, i.e. $N=2$ theory
has 16 real supercharges.} 
supersymmetric Yang-Mills-Higgs theory 
with gauge group $SU(N)$ broken down to the maximal torus $U(1)^{(N-1)}$ admits a new type of nonsingular instanton
solution. This instanton, despite the broken gauge group and broken
scale invariance of the theory is not singular due to
the fact that it  carries electric charge under the unbroken 
Abelian subgroup. This charge supports the instanton against
collapse. Since the instanton carries both topological charge $k$
(winding number)
and electric charge $q$, the authors of \cite{LT} have christened it a {\it
dyonic} instanton. It was shown in \cite{LT} that this configuration
is supersymmetric, preserving 1/4 of the supersymmetries.
We will show that this solution solves the full
non-abelian DBI action for the $D4$ brane. 
We interpret this solution, for
the case of a $SU(2)$ gauge group, as the superposition of
a $D0$ brane with a fundamental string connecting two parallel $D4$ branes. The
geometry of this configuration is interesting and new compared to the
shape of the brane with monopole solutions given in \cite{hashi}. Due
to the presence of the  delocalised source of the electric field on
the brane, a  new effect of blowing-up of the brane appears. We also
comment on the problem that one encounters when trying to abelianise the solution.

\vskip20pt
\section{Systematics of the NDBI calculus}

Before discussing the nonabelian dyonic instantons, we will now first
present a systematic method for closed-form calculations involving the
NDBI action.
As proposed by Tseytlin \cite{tsey} the NBI action for the $D9$ brane is given by
\bea
\label{NDBI9}
L_{ NBI} (F) 
   = Str \sqrt{- det(\eta_{\mu\nu} +  T^{-1}  F_{\mu\nu} ) } \ .  
\eea
where $\mu,\nu=0,...,9$ and $T={1\over 2\pi \alpha'}$.
The symbol Str in (\ref{NDBI9}) is defined in the following way. By expanding
the {\it abelian} BI action as 
\bea
\label{expBI}
\sqrt{-det(\eta_{\mu\nu}  + F_{\mu\nu})}  = \sum_{k=0}^{\infty} C^{\mu_1\nu_1... \mu_{2k} \nu_{2k}}
 F_{\mu_1\nu_1} ... F_{\mu_{2k} \nu_{2k}} \ . 
\eea
one defines the Lorentz tensors $C^{\mu_1\nu_1... \mu_{2k} \nu_{2k}} $. Then
the NBI action is {\it defined} as a power series

\bea
\label{expNBI}
Str \sqrt{-det (\eta_{\mu\nu} I  + F_{\mu\nu})}  \equiv
 \sum_{k=0}^{\infty} d_{a_1 ...a_{2k}} C^{\mu_1\nu_1... \mu_{2k} \nu_{2k}}   
 F^{a_1}_{\mu_1\nu_1} ... F^{a_{2k}}_{\mu_{2k} \nu_{2k}} \ .  
\eea
Here the tensors
\bea
\label{dtensor}
D_{a_1 ... a_p} &=& {1\over n!} (T_{a_1}...T_{a_p}+all \ permutations ) \nn  
d_{a_1 ... a_p} &=& tr(D_{a_1 ... a_p}) 
\eea
are totally symmetric, adjoint action invariant tensors of the gauge group.\footnote{A major achievement of this action is that it reproduces
correctly all terms, up to the $\alpha^{2} F^{4}$, from 
string theory calculations. }
Strictly speaking, given this definition of Str,  one can 
only work with a power expansion form of the action. However,
extending the Tseytlin definition as follows allows one to do
calculations under the Str without writing out a power expansion.
Following definition (\ref{expNBI}), one first  introduces a new type of
object: a function of the  the Lie algebra variables, $\Hf$.

Let $f = f(A_{1}, ... ,A_{n})$ be a  $C^{\infty}$  function  with the expansion

\bea
\label{abelf}
f = \sum_{k=0 \atop {h^{k}_{1}+...+h^{k}_{n}=k} }^{\infty} f^{h^{k}_{1}....h^{k}_{n}}
A_{1}^{h^{k}_{1}}... A_{n}^{h^{k}_{n}} \ . 
\eea
Then the corresponding function $\Hf$ over Lie algebra variables $\HA_{i}=\HA_{i}^{a}T^{a}$,
$i=1,...,n$ is defined as
\bea
\label{ndbif}
\Hf \equiv  \sum_{ k=0 \atop h^{k}_{1}+ \cdots +h^{k}_{n}=k}
^{\infty} f^{h^{k}_{1}....h^{k}_{n}}
\HA_{1}^{\bf{a_{1}}}  \cdots  \HA_{n}^{\bf{a_{k}}} D_{\bf{a_{1}}...\bf{a_{k}}} 
\eea
where a $D$-tensor is defined as in (\ref{dtensor}).

Next, we define operations with hat functions.
Additivity and multiplication by a  scalar is naturally induced from the
algebra of the ``abelian'' functions $C^{\infty}$. Note that the induced
addition for the hat functions is commutative: $\Hf \ + \  \Hg =
\Hg \ + \ \Hf$. 
Multiplication in $C^{\infty}$ induces a natural
multiplication among the hat functions as follows.
For simplicity we will talk below about  functions of a
single Lie algebra variable, and comment on the 
generalisation to the case of several variables only where it has specific features.
For functions $\Hg(A)$ and $\Hl(A)$ the first guess for the induced
product $\Hg(A) \Hl(A)$ might be 

\bea
\label{prod1}
\Hg(\HA) \Hl(\HA) \equiv \sum_{k=0}^{\infty}  \sum_{m=0}^{\infty} 
g^{{k}} l^{{m}} 
\ \HA^{\bf{a_{1}}} ...\HA^{\bf{a_{{k}}}}
\HA^{\bf{b_{1}}}...\HA^{\bf{b_{{m}}}} \
D_{\bf{a_{1}}...\bf{a_{k}}} \ D_{\bf{b_{1}}...\bf{b_{m}}} 
\eea 
However, this is not a well defined operation since the product of two hat
functions is not a hat function.
If the {\it abelian} counterpart functions satisfy $f=l g$, then
this product does not imply $\Hf=\Hl \ \Hg$ since 
\bea
D_{\bf{a_{1}}...\bf{a_{k}}}D_{\bf{b_{1}}...\bf{b_{m}}} \neq
D_{( \bf{a_{1}}...\bf{a_{k}}} D_{{\bf b}_{1}...{\bf b}_{m} )}
\eea
Only in the case of a  {\it single } variable $A$ does the equality $\Hf
= \Hl \Hg$ hold, since
$\HA^{\bf{a_{1}}}...\HA^{\bf{a_{{k}}}} \HA^{\bf{b_{1}}}...\HA^{\bf{b_{{m}}}}$
is a totally symmetric expression, and hence induces symmetrisation in
the product of $D$-tensors. 
The approach adopted in $\cite{hashi,wash}$ was to treat all
hat functions under the Str as {\it abelian}. This means that the product
(\ref{prod1}) is modified to 
\bea
\label{prod2}
\Hg(\HA)*\Hl(\HA) \equiv \sum_{k=0}^{\infty}  \sum_{m=0}^{\infty} 
g^{k} l^{m} 
\HA^{\bf{a_{1}}}...\HA^{\bf{a_{{k}}}} \HA^{\bf{b_{1}}}...\HA^{\bf{b_{{m}}}}
D_{(\bf{a_{1}}...\bf{a_{k}}} D_{ \bf{b_{1}}...\bf{b_{m})}} 
\eea 
where we have  introduced additional symmetrisation over products of
$D$-tensors. This product is obviously closed and moreover  respects
associativity and is commutative. 
When supplied with operations for  scalar multiplication,
addition and multiplication,  the set of  hat functions forms an algebra.

Another property that we will use is that
variations of the hat functions can be taken  
as in the abelian case, i.e. one can take variations without worrying
about  the  gauge group structure (as long as we are assuming the
'~*~' product)
\bea
\delta \Hf &=&  \sum_{k=0}^{\infty}  
\ k \  g^{{k}}\ \HA^{\bf{a_{1}}}...\delta \HA^{\bf{a_{{k}}}}
D_{\bf{a_{1}}...\bf{a_{k}}} \\ 
&=& \sum_{k=0}^{\infty}  
\ k \  g^{{k}}\ \HA^{\bf{a_{1}}}... \HA^{\bf{a_{{k-1}}}}
D_{\bf{a_{1}}...\bf{a_{k-1}}}* \delta \hat A \\
&\equiv& {\delta \Hf \over  {\delta \hat{A}} } * \delta \hat{A}
\eea
Hence the variation of the action (\ref{NDBI9}) is  
\be
\delta L = Str[\sqrt{-det \hat{H_{\xi\rho} }} * (\hat{H}^{-1})^{ \mu\nu} *
\delta F_{\mu\nu} ] \ ,
\ee
where
$\hat{H}_{\mu\nu} \equiv \eta_{\mu\nu} +  F_{\mu\nu}$ and $\delta F_{\mu\nu}=D_{\mu}\delta A_{\nu} - D_{\nu} \delta A_{\mu}$ . Since for $F$ (and quantities
derived from it by T-duality) there is no distinction between 
ordinary and hat functions, we will henceforth write them without
the hat.

All we have done  up to now  referred only to the $D9$
brane. To get the NDBI action for any $Dp$-brane, one goes to the
static gauge, identifies world volume
coordinates (along which we want to T-dualise) with background
coordinates and applies T-duality 9-$p$ times to (\ref{NDBI9}). 
After T-duality  we get the action 
   
\bea
\label{NDBI}
S_{p} &=& T_p \int d^{p+1} x \
 Str \sqrt{- det} 
(\eta_{\mu\nu}+   T^{-1} F_{\mu\nu}) \nn
   &=& \ 
  T_p   \int d^{p+1} x \
 Str  \bigg{[}  
    \sqrt {-det \hat{{\cal G}}_{mn}} * \sqrt { det \hat{{\cal G}}_{rs}} \bigg{]} \ , 
\eea

\bea
\hat{{\cal G}}_{mn} \equiv  \eta_{mn} + \hat{{\cal G}}^{rs}(X)  D_m X_r D_n X_s  
+ T^{-1} F_{mn} \ , \ \ \ \ \ 
\hat{{\cal G}}_{rs}(X)  \equiv \delta_{rs} - i T [X_r, X_s] \ ,
\eea
where indices $m,n=0,...,p \ $ and $r,s,=p+1,...,9$.
It is important to note that all hat functions under the Str
are functions of the variables $F$ and quantities derived from it by
T-duality,  
\bea
\label{notation}
F_{rs} &=& -iT^2 [X_r,X_s]  \ , \nn   
F_{mr} &=& T D_m X_r= T(\partial_m X_r - i [A_m,X_r]) \ ,  \nn
F_{mn}(A)&=& \partial_m A_n - \partial_n A_m - i [ A_m, A_n] \ .
\eea
One is $\bf{not}$ associating Lie algebra generators to $X$'s and
$A$'s, but just to $F$'s under Str. Hat functions are by definition functions
of Lie algebra variables $F$, not $A$ or  $X$. As soon as  we pass from $F$'s to  $A$'s or $X$'s we 
lose the ability to write expansions in closed hat function form.  
In  order to  go from the first to the second equality in (\ref{NDBI}) we have
used
\vskip10pt
\begin{eqnarray}
det \left( \ba{cc} A & B \\ C & D \ea \right)  
&=& det(D) det(A-BD^{-1}C)
\end{eqnarray}
where here it is understood that $det$ is the hat determinant.
If we compare the action (\ref{NDBI}) with the abelian case of the appropriate
dimension we see that  the nonabelian action for $Dp$-branes is not just
the abelian action with the  replacement $D \leftrightarrow \partial$ and
Str introduced.
There are additional commutation terms $[X,X]$  and $[A,X]$ in the nonabelian
action. 

Another operation that one often performs under $Str$ is
\bea
\label{part}
tr( \Hf * \delta F_{mn})= tr\bigg{(} \Hf  (D_{m} \delta A_{n} - D_{n}
\delta A_{m}) \bigg{)}.
\eea
Here in the second equation we have passed from the ' * ' product to the 
ordinary product. We can do this  due to the fact that $F$ carries only
one Lie algebra index. In each term of the Str  we can always
cyclically permute the $T^{a}$ generator (corresponding to the $F$) under the trace to the right. So this operation of
passing from ' * ' to the ordinary product is possible {\it only}
under the Str and only for the expressions of the form $tr(f*l)$ where
$f=f^a T^a$ and $l$ arbitrary.  

Next we want to define the {\it covariant derivative} of the hat function. 
In order to do this we  first note that any hat function $\Hf$, valued in
the $SU(N)$ gauge group, can be written as 
\bea\label{foot}
\Hf =f^{a} T^{a} + f^{0} I 
\eea
where $I$ is the identity
matrix. This follows from the property of the $SU(N)$  generators
$T^{a} T^{b} = {c \over 2} \delta^{ab} + {1 \over 2} d^{ab}_{c} T^{c}  + {i
\over 2} f^{ab}_{c} T^{c}$. Hence, any product of generators can be
expressed as linear combination of the identity matrix and
the generators. The covariant derivative is then defined as 
\bea
\label{hatD}
D_{m} \Hf \equiv \partial_{m} f^{a} T^{a} + \partial_m f^{0} + f^{a}
[A_{m},T^{a}] 
\eea
Although it  is not obvious from the definition (\ref{foot}) that
the covariant derivative of a hat function is itself  a hat function, one can prove
this easily by writing  $\Hf$ explicitly  in expanded
form. However, the abelian counterpart function of   $D\Hf$ is
not obvious. Generically, $D\Hf$ is of the form: $D\Hf= \partial \Hf +
\hat{H}$ where $\hat{H}$ is not directly expressible in terms of 
$\Hf$. This is in contrast to ordinary nonabelian gauge theory
where the analogue of $\hat{H}$ is a commutator of a gauge potential and a
function (here $\hat{H} \neq A * \Hf - \Hf * A$). In order to calculate
 $\hat{H}$ explicitly one has to use the full
expanded form. As long as we treat  $D\Hf$ as a single
object (not worrying about its explicit form) we can treat all
expressions as abelian ($D\Hf$ is a hat
function). However, if we want to perform some calculation which
depends explicitly on the structure of
$D\Hf$ (for example, to calculate $\delta D \Hf$) we must work 
with the definition of $D\Hf$ in expanded form.

Now we can do a partial integration of (\ref{part}) so that it
becomes
\bea
tr( \Hf * \delta F_{mn}) = tr ( (D_{m} \Hf) \delta A_{n} - (D_{n} \Hf)
\delta A_{m} ) \ .
\eea

In summary, as long as we are dealing with hat
functions under Str as functions of the Lie algebra variable $F$ and
perform operations which are well defined for them, i.e. 
operations which send hat functions into hat functions 
(multiplication by a  scalar, addition, multiplication and partial differentiation)
we can treat all expressions under the symmetrised trace as
abelian. On the other hand, the change of variables from $F$ to $A$
and explicit calculation with covariant derivative of hat function is possible only in the expansion form.

\vskip20pt
\section {Dyonic instanton as solution of the nonabelian DBI action}
 
In a recent paper by Lambert and Tong  \cite{LT}, it was shown that Yang-Mills-Higgs theory  broken to the $U(1)$ subgroup admits a new type of the instanton solution, the dyonic instanton. This instanton is  a BPS instanton embedded in the 4+1, $SU(2)$ Yang-Mills-Higgs theory broken to $U(1)$. This instanton  is charged with 
respect to the unbroken $U(1)$ gauge group. 
We will now apply the rules for the hat functions in order to show
that the dyonic instanton is a solution to the NDBI action.

Since the dyonic instanton 
is characterised by only one transverse scalar being nonzero, the 
action (\ref{NDBI}) in this case simplifies greatly---all commutators vanish
identically. In what follows all functions are understood to be hat
functions and all products ' * ' products, but to simplify notation we will not write them explicitly.
 
In order to show that this configuration satisfies the equations of motion
of the NDBI action it is convenient to first rewrite the action (\ref{NDBI})
as the non-abelian BI action. Using the T-duality rules (\ref{notation}), we can
rewrite the NDBI action for the D4 brane as \cite{hashi}
\bea
\label{Faction}
L = {1\over{(2 \pi \sqrt{\alpha'})^{p+1}g_{s}}}{Str} \sqrt{-{det} (\eta_{mn} I  + T^{-1}F_{mn})} 
\eea
where $m,n=0,...,5$ and  $X$ is in the direction one and $i,j=2,...,5$
correspond to the spatial directions of the D4 brane. In action
(\ref{Faction}) we have we have reinstated the brane tension
(i.e. string coupling and factors of $\alpha'$). By expanding the DBI
action and comparing with the Yang-Mills-Higgs actions one finds  that
the Yang-Mills coupling is related to the string parameters as
\bea
{1\over g_{YM}^2} =
{T^{{p+1}\over2}\over{(2\pi\sqrt{\alpha'})^{p+1}g_{s}}}  \ .
\eea
Varying (\ref{Faction}) with respect to $A$  gives 
\bea
\label{vari}
\delta L = {1\over{g_{YM}^{2}}}{Str} \bigg{(} \sqrt{-{det} (T \cdot I + F )}
 [(T \cdot I + F)^{-1}]^{ mn} ( D_m \delta A_{n} - D_{n} \delta A_{m} )  \bigg{)} \ .
\eea

Minimising the expression for the energy density of the
Yang-Mills-Higgs langrangian, the authors of \cite{LT} have found the BPS equations
for the dyonic instanton to be 
\bea
\label{BPS}
E_{i} =TD_{i}X\nn
F_{ij}= \mp \tilde{F}_{ij} \ .
\eea
with $i=2...5$ and $E_{i} \equiv F_{0i}$.
In addition to the BPS equations one has to satisfy the Gauss law
\bea
\label{gauss}
D_{i}E^{i} = T^{2}[X,D_{0}X] \ .
\eea
After performing the variation, we are allowed to plug the BPS equations
(\ref{BPS}) into (\ref{vari}). For the BPS equations the square root in
(\ref{vari}) linearises:
\be
{det} (I + F) = { \bigg{(} 1-{1\over 4} {tr}F^{2} \bigg{)}}^{2}
\ee
where ${tr}$ denotes the trace over the Lorentz indices.
Going to the basis in which $F$ is
skew-diagonal and evaluating the inverse of $(I +
 F)^{-1}$ for the BPS configuration, one gets
\bea
\label{F}
(I + F)^{-1} = {1 \over (1- {1\over4} trF^{2})}  \left(
\begin{array}{cccccc}
-(1- {1\over4} tr F^{2}) & 0  & 0 & 0  & 0 & 0 \\
0   &  (1- {1\over4} tr F^{2}) & 0 & 0  & 0 & 0 \\
0  & 0  & 1 & f  & 0 & 0 \\
0  & 0  & -f & 1  & 0 & 0  \\
0  & 0  & 0 & 0  & 1 & f  \\
0  & 0  & 0 & 0 & -f & 1 \\
\end{array}
\right) \ ,
\eea
where $f^2 = {1\over4} {tr} F^{2} $. 
So we can rewrite (\ref{vari}) as 
\bea
\label{vari1}
\delta L = {Str} \bigg{(} F^{ij} ( D_i \delta A_{j} - D_{j} \delta A_{i} )  \bigg{)} \ .
\eea
Since this expression now only involves two factors, there is no need
to distinguish between ${Str}$ and ordinary ${tr}$. One can now
apply ordinary integration by parts to obtain the equation of motion
\footnote{Note that the equations of motion for $X$ and $A_{0}$ are
automatically satisfied when we  plug the BPS conditions into
(\ref{vari}). }

\bea
D_{i}F^{ij} = 0 \ ,
\eea
which is satisfied for the self-dual configurations due to the Bianchi identity.

The energy of the system is obtained by performing a Legendre
transformation of the Lagrangian (\ref{NDBI}).
Starting  from the Lagrangian   (\ref{NDBI}) and expanding  the
determinant under the square root one gets
\bea
\label{sqrt}
L &=& {1 \over g_{YM}^2}{Str} [1 + T^2 DX.DX - E^{2} + T (E.DX)^{2} - T^2 (DX.DX)E^{2} - {1\over2}
F^{2} + T^2 DX.F^{2}.DX - E.F^{2}.E \nn
&-& {T^2 \over 2} (DX.DX) {tr}F^{2} +
{1\over2} E^{2} {tr}F^{2} {1\over4} {tr}F^{4} + {1\over8}
( {tr}F^{2})^{2}]^{1\over2} \ .
\eea
Next apply the Legendre transformation
\bea
{\cal E} = {STr}({{\pa L} \over {\pa E}}.E) - L \ .
\eea
The energy of the
system is  
\bea
\label{ener}
{\cal E} = {1\over g_{YM}^2} Str \bigg{(} (1 &+& T^2 DX.DX   - {1\over2}
F^{2} + T^2 DX.F^{2}.DX  - {T^2 \over 2} (DX.DX) {tr}F^{2}  \nn
 &+& {1\over4} {tr}F^{4} +
{1\over8} ( {tr}F^{2})^{2}) {det(....)}^{-1\over2} \bigg{)}
\eea
where here $det(....)$ is short for the expression under the square root
of (\ref{sqrt}). The denominator in (\ref{ener}) should be understood as the
inverse of the hat  function  with respect to the star product.  
For the BPS configuration the energy density becomes 
\bea\label{energ}
{\cal E} = {1\over g_{YM}^2}Str \bigg{(} 1+E^{2}-{1\over 4} trF^{2} \bigg{)} \ .
\eea
This is the same expression as was obtained in
\cite{LT} by BPS arguments to minimize the energy. Note however, that
the solution that we get is not obtained by minimizing the
energy, but by directly showing that the BPS equations of \cite{LT} solve
the full equations of motion. 
 
The second of the BPS equations (\ref{BPS}) is
the ordinary (anti)self duality equation for the nonabelian instanton.
The solution in the singular gauge \cite{bela} is given by 
\bea
A_{i} = 2{{\rho^2} \over {x^{2} ( x^{2}+ \rho^{2})}}
\eta^{a}_{ij}x_{j} {{\sigma^{a}}\over2} \ ,
\eea
where $\eta^{a}$ are the self-dual 't Hooft matrices,  $\sigma^{a}$ are
the SU(2) generators, and the parameter $\rho$ is the size of the instanton.
For the static field configuration ($\partial_{0}=0$), after setting
$A_{0}=-X$, and after use of the Gauss law (\ref{gauss}), the  second BPS
equation reduces to  the covariant Laplace equation in the background of instanton \cite{LT} 
\bea
D^2 X =0 \ .
\eea
This equation has a unique solution  \cite{dore} for each $v$-VEV of
the scalar field X. It is given by 
\be
\label{X}
X = {g_{YM}\over T} v  { x^2 \over {(x^2+\rho^2)}} { \sigma^{3} \over 2 } \ .
\ee
When the BPS equation are satisfied, the field configuration has a mass
which is equal to the sum of the charges
\bea
\label{charg}
k &=& {1\over g_{YM}^2} tr \int_{WS} d^{4}x F_{ij}F^{ij}   \ , \\
q &=& {T \over v {g_{YM}^2}} tr \int_{WS} d^{4}x E DX = {T \over v {g_{YM}^2}} tr \int_{S^{3}_{\infty}}
d \vec{\sigma} . \vec{E} X = 4{\pi^2} {\rho^2} v \ ,
\eea
where $WS$ stands for world space of the D4 brane, and in the second equation when
going to the surface integral we have used Gauss' law. Charge $k$ is 
proportional to the instanton number, while $Q=qv$ is total electric charge carried by the
instanton. Since $Q$ is a conserved
charge 
we see  that instanton
stabilises at size $ \rho = {Q^{1/2} \over {2 \pi v}}
$ \cite{LT} which is determined by the electric charge of
the instanton and the VEV of the scalar field. Note that the electric
charge (i.e. the electric part of the energy) is independent of the
string coupling while the magnetic part depends on the Yang-Mills
coupling as $1\over g_{YM}^2$ i.e. as $1\over g_s$. This suggests that
in the brane picture the electric part of the configuration originates
from the fundamental brane (string) while the magnetic part
corresponds to the D-brane. That this is indeed the case will be shown
in the next section.

To conclude this section we shall comment briefly on the abelian
solution. One can easily derive  the {\it abelian} BPS
equations by minimizing the expression for the
energy. They are the same as (\ref{BPS}) up to a replacement $D
\leftrightarrow \partial$. Because of this replacement, the  equations for $E$ and $F$ are no longer coupled, as 
in the non-abelian
case. 
The equation for $E$, after use of Gauss' law, implies 
that $X$ is a harmonic function on the worldspace ($X= q/4\pi r^{2}$),
while the selfduality
equation has the solution 
 \be
A_{a} = \sum_{r=1}^{r=3} (I^{r})_{ab} \delta ^{bc} \pa_{c} V_{r}
\ee
where the $I^{r} $  are three independent, anti-self-dual, complex
structures in $\bC^2$ obeying the quaternionic algebra  relations and $V$ is
harmonic on the world space $\bC^2$ ($V={K\over{r^2}}$). Both the solution for $X$ and
the solution for $A$ are
{\it singular} and of {\it infinite} energy.
Following the logic of \cite{paper,koch} one can try to get  a { \it non-singular, finite
energy} solution by embedding the brane in the appropriate background.
Presence of the brane background  \cite{paper} induces a ``cut-off'' in the
distances which one can probe on the worldvolume of the $D4$ brane, leading to a finite energy of the
electric part of the solution ($E_{electric} = {T \over g_{YM}^2} \int d^4 \sigma \
{\vec E} \cdot {\vec \pa} X = qv $). However, the part of the energy corresponding to the
instanton behaves as   
${T\over g_{YM}^2} \int_{\bR^4} d \sigma^{4} F_{ab} \tilde{F}_{ab} \sim
\oint_{S^{3}_{r=r_{0}}} A \pa A \sim ({v\over q})^{2} K^{2} $.
We see that the magnetic part of the solution depends on the
separation of the brane from the background (v) and moreover that this
dependence is quadratic. We have not yet  been able to understand the cause
of this behaviour of the abelianised solution.

\vskip20pt
\section{ Brane interpretation of the solution}

It is known that the instanton solution in the gauge theory living on the
stack of $D4$ branes has an interpretation as a $D0$ brane within the $D4$ brane
\cite{doug,witt}. A large instanton is interpreted as a $D0$ brane which is dissolved
into the flux of the YM gauge field,
while  a singular instanton corresponds to an undissolved $D0$ brane within the $D4$ brane. 

The reasons for this interpretation are  the following.  
The number of  supersymmetries in the field theory preserved by the
instanton solution is 1/2 of the supersymmetries of the brane theory. This is the same as the number of supersymmetries preserved by the $D0-D4$ system. Furthermore, the instanton solution carries the same RR charge
as a $D0$ brane. This can be seen by looking at the Chern-Simons (CS)
term in the DBI action of the $D4$ brane. In a constant $C$-field, the CS  term with 
selfdual gauge field $F$ reduces to 
\bea
{1\over 2} T^{-2} T_{4} \int C_{1} \wedge Tr( F \wedge F) = T_{0}\int
C_{1} \ ,
\eea
since $ \int Tr( F \wedge F) = 8 \pi^{2}$ and $T_{0}=(2 \pi
\sqrt{\alpha'})^{4} T_{4}$. So we recover the coupling of the $D0$ brane
to the RR field.

Finally, the moduli of the effective  action describing a system of two
$D4$ branes and one $D0$ brane in the
Higgs branch  are the same as
the moduli of the $SU(2)$ instanton. To illustrate this fact lets
consider the case of the $k=1$, $SU(2)$ instanton. The number of
moduli of this instanton is eight: four corresponding to the position
of the instanton, three corresponding to the orientation of the
instanton in the $SU(2)$ group (i.e. the  embedding of the $U(1)$ in the $SU(2)$ group) and one corresponding to the instanton size. 

On the other side, to count the number of the moduli in the $D0-D4$ system we have to look at the  potential term in the effective action
for the $D0-D4$ system \cite{polc}  
\bea
\label{Leff}
{g_{D0}^{2} \over 4} \sum_{A=1}^{3} ( \chi_{ia}^{ \dagger} \sigma_{ij}^{A} \chi_{ja}
)^{2}  +  \sum_{i=1}^{5}(X_{i}-X'_{i})^{2} \chi^{ \dagger} \chi \ .
\eea
Here  $\chi_{i}$ is a doublet of complex  hypermultiplet scalars,
the  index $a=1,2$
labels two D4 branes and,  
$X_{i}$ and $X_{i}'$ are scalars in the vector multiplet.

In the Higgs
branch, the $D0$ brane is within the $D4$ brane ($X_{i}= X_{i}'$), so the
second term in (\ref{Leff}) vanishes
identically. Each of the remaining three terms in
(\ref{Leff}) has to vanish separately, and so we have three constraints
on the field $\chi$. Moreover, using the $SU(2)$ invariance of  
(\ref{Leff}) we can fix one component of $\chi$. There are eight real
hypermultiplet scalars in $\chi_{a}$. After use of the constraints and $SU(2)$ invariance  one is left with four moduli. These moduli correspond to moduli describing the embedding of the instanton and instanton size. The additional four moduli come from the $D0-D0$ strings and describe the position of the $D0$ brane within the $D4$ brane. 
   
The instanton solution in our case is
unmodified compared to the un-Higgsed YM theory, except that the size of
the instanton is no longer a free modulus, but it is fixed  by the
separation of the branes and by the
electric  charge which the instanton carries. Hence, we can
interpret our instanton  as a $D0$ brane dissolved in the YM flux in the worldvolumes of two separated $D4$
branes. 

However, we still have to explain the origin  of the electric
field on the D4 brane. A useful observation  is that the  electric part of the energy
of the system is {\it independent} of the string coupling. This is
a know property of the energy of fundamental objects, such as  $F1$
strings. This is in contrast to the magnetic part of the
energy which is proportional to  ${1\over g_s}$, i.e. it behaves as the
mass of the D-brane.

Additional information is gained from the
effective geometry of the configuration. So let us look more closely at the shape of the
brane for our field configuration. 

\begin{figure}
\centerline{\psfig{file=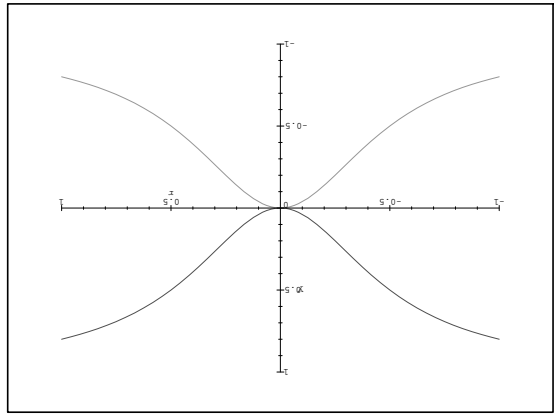,angle=90,width=8cm}}
\centerline{\psfig{file=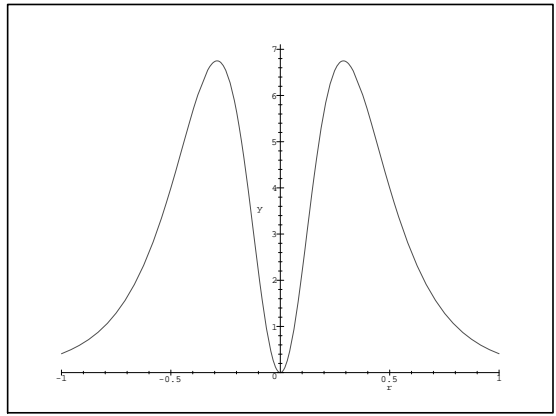,angle=270,width=8cm}}
\caption{(a) shape of the branes with the dyonic instanton (upper and
lower branches of the picture correspond to the first and
the second $D4$ brane); $\  $ (b) energy density of the electric
field on the worldvolume of the brane.}
\end{figure}

Along the lines of
\cite{hashi} one can immediately read off  the shape of
the brane  from (\ref{X}). This is depicted in the fig 1.a. Far away from the core of the instanton, the shape of the brane looks like the brane pulled by  fundamental
strings, i.e.  $X$ behaves as $ X \sim {v\over2}(1- {\rho^2 \over r^2})
$. However, at  $r_{0} = {\rho\over\sqrt{3}}$ the brane changes
shape from concave to convex. Also, if one calculates the
energy density corresponding to the electric field, i.e. the intensity
of the electric field (see fig 1.b), one finds the expression

\be
{\cal E}_{E} = E_{i}^2 \sim {r^2\over(r^2+\rho^2)^4} \ .
\ee
 
This expression has a maximum at distance  
$r_{0}= {\rho \over\sqrt{3}}$, at  the same position at which  the 
shape of the brane changes from concave to convex. Hence, it seems like the source of
the electric field on the brane is located on the sphere of radius          $r_{0}$.
To determine the origin of the electric field we first recall that 
the D-brane is perturbatively defined in terms of $F1$ strings with
Chan-Paton (CP) factors attached to the end
of the string. CP factors are
sources of the gauge group on the brane and different brane
configurations in the effective description correspond to different
gauge configurations. For example, an $F1$ string ending on the $D3$ brane  
in the effective description is  an electric charge in the
world volume gauge theory of the $D3$ brane \cite{mald}, and it turns
on the electric field in the worldvolume of the brane. On the other
hand a $D1$ string turns on the magnetic components of the gauge field
in the worldvolume of the brane, and is represented as a magnetic
monopole in the worldvolume
gauge theory.

When the $F1$ (or ($p,q$) string) terminates on the brane, the effective geometry
of the brane is such that  it  bends under the tension of the
string. Due to the fact that the brane has a tendency to
minimize its surface area, the brane has  a convex form  when pulled
by the string. In fig 2. the shape of two separated branes, pulled by
the fundamental string which connects them, is shown. For the
configuration shown in fig 2. the electric field on the branes grows
as we approach the point at which the string is attached. 
\begin{figure}
\centerline{\psfig{file=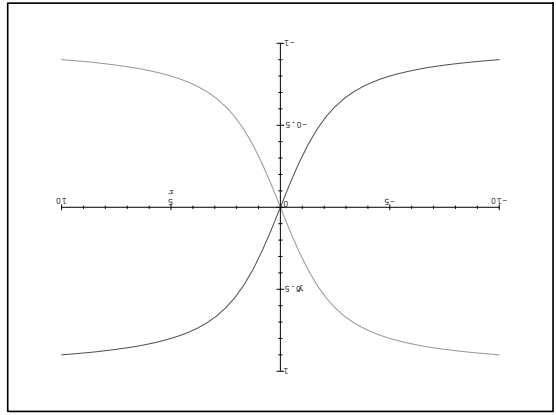,angle=90,width=8cm}}
\caption{shape of the branes with the monopole}
\end{figure}

In our case however, the effective  geometry is different since the source of the electric
field is not localised at a  point.

Far from the core of the instanton, the  electric
field falls-off as $1\over x^{3}$. This  is exactly the  behaviour of
the electric charge in 4+1 dimensions. So, far away from the core
of the instanton, the geometry of the brane is the same as in the previous
case where the $F1$ pulled the  brane. 
All this suggests that the correct  interpretation of the electric field on the brane is that 
it originates from  the $F1$ string which connects  $D0$
and $D4$ branes. The effective description of a $D0$ brane within a
$D4$ brane is that of the instanton: in other words  when the $D0$
brane is in the $D4$ brane it dissolves into
the flux. Hence, since the $F1$ string terminates on the $D0$, the same
thing happens to the $F1$ string: it dissolves into the electric flux on the brane.
As we start to separate the branes, the instanton starts
to shrink, but its complete collapse is halted by the electrostatic
repulsion (which acts as negative tension on the brane surface). 
The presence of the delocalised  electric source causes the ``blowing up'' of
the brane surface which counterbalances the  brane which tends to collapse. 
\footnote{This is to some  extend similar to the Dirac model of the
electron  as a charged bubble, which is prevented from collapse
by electric repulsion, 
except that in our case the electric field is confined to live just on the surface
of the brane and not in the full spacetime.} 
As we increase the separation between the branes,
the force that pulls the brane, and hence the $D0$ within it, increases 
\footnote{The tension of the string connecting the separating branes is
constant, but the total force acting on the branes is $\sim v T$,
so it increases as we separate the branes.}
and
hence the radius at which the instanton stabilises becomes smaller.
 
In summary, if we start with the system of coinciding $D4$ branes and a $D0$
brane within  them, then the presence of the $D0$ brane in the effective
theory is seen  as  either a ``large'' $\rho$ instanton
 (a dissolved $D0$ brane, which is described by the Higgs branch of the effective theory) or a  zero size instanton. 
The singular instanton is interpreted as the $D0$
brane that  can be pulled out from the stack of $D4$ branes. Once we
separate the $D0$ brane from the stack of the $D4$ branes, the system  is described by the Coulomb branch  of the full effective action
(\ref{Leff}). 
However, if we  separate the stack of $D4$ branes with the
``large'' $D0$ brane (instanton) sitting within them, then the $D0$ brane will start
shrinking and it will stabilise at  a radius which is
inversely proportional  to the separation between the branes. 
\footnote{In the previous discussion we were talking about BPS
configurations, in which  kinetic energy of the system was always zero. Hence,
when we talk about the process of separating the branes, or
shrinking of the instanton, we assume a quasi-static process between
two BPS configurations.}   

It is interesting to note 
\footnote{I thank P.K.Townsend for pointing out this to me.}
that for each dyonic instanton (characterised
by charge $Q=qv$ and instanton number $k$), there is a
corresponding configuration of a string of length $v$ and separated
set of 
$k \ D0$ branes which has the same energy as the dyonic instanton. However
this state cannot be described in the field theory living on the $D4$
brane. Although these two
configurations have the same energy, it seems likely that the regions of the full
moduli space, corresponding to the dyonic instanton and the $D0-F1$
system, are separated by a potential barrier. In support of this, we
 note that to ``pull out'' a $D0$
brane one first has to shrink it to zero size, and that requires 
infinite separation of the $D4$ branes. 
However, to describe the full moduli space one would need to analyse
the effective action of the full $D0-D4$ system which we leave for 
future investigation.

\section{Conclusions and  discussion}

In the first part of the paper we have tried to present systematic
rules that one should use when dealing with the nonabelian DBI 
action. These rules allows one to treat the NDBI action as abelian, and
work in closed form (with all terms in the expansion). However, we
have pointed out that there are limitations to this approach, i.e that
after certain operations (like taking covariant derivatives of hat
functions) one loses the ability  to further express things in a
obvious closed form. 
As for future directions, it would be interesting to try to apply
this approach to deduce the Hamiltonian formulation of the NDBI
action. In this case one is facing the new problem 
of solving the nonabelian constraints. We hope to address this question
in future work.

In the second part  we have seen a new effect in the
effective brane geometry--- the blowing up of the brane. It would  we
interesting to try to use  this effect to construct compact
branes. Up to now all compact brane configurations were constructed  either by wrapping branes on a compact
submanifold of  the background (in this case the brane is prevented from 
collapse by the background geometry) or by putting the brane  in
an external RR-field \cite{myer}. It would be interesting to explore the possibility of having compact branes due to a nontrivial gauge configuration on the worldvolume of the brane.

\section*{Acknowledgements} 
I would like to thank Dominic Brecher, Neil Constable, David Tong, Ian
Hawke, Eduardo Eyras, Nick Manton, Jonathan Moore, Malcolm Perry, Andy Neitzke, Adam Ritz for 
discussions. I am especially grateful to Paul Townsend and Kasper
Peeters for many useful discussions, a lot of patience and great moral support.

\bigskip

\end{document}